\documentclass{iaus}
\usepackage{graphicx}
\usepackage[version=3]{mhchem}
\title[Cold and warm protostellar structure]{The cold and warm physico-chemical structure of embedded protostars.}
\author[N. M. Murillo]{N. M. Murillo} 

\affiliation{Leiden Observatory, Leiden University, P.O. Box 9513, 2300 RA, Leiden, the Netherlands \\ email: {\tt nmurillo@strw.leidenuniv.nl}}

\pubyear{2017} 
\volume{332} 
\pagerange{1--12}
\jname{Astrochemistry VII -- Through the Cosmos from Galaxies to Planets}
\editors{M. Cunningham, T. Millar \& Y. Aikawa, eds.}
\begin{document}

\maketitle

\begin{abstract}
The physical evolution of low-mass protostars is relatively well-established, however, there are many open questions on the chemical structure of protostars.
The chemical fingerprint generated in the early embedded phase of star formation may be transmitted to the later stages of star, planet and comet formation.
The factors that influence the chemical fingerprint are then of interest to study, and determine whether the chemical structure is inherited from the parent cloud or product of the physical processes during star formation.
Results of observations and modelling of molecules that trace the cold and warm extended structures of embedded protostars are briefly presented here.
Two multiple protostellar systems are studied, IRAS 16293-2422 and VLA 1623-2417, both located in $\rho$ Ophiuchus.
We find that the physical structure of the protostars, that is the disk(-like) strucutres, outflow cavity and different luminosities, are important factors in determining the chemical structure of these embedded protostars.
\end{abstract}

\firstsection 
\section{Introduction and observations}

Two low-mass embedded Class 0 protostars, both located in $\rho$ Ophiuchus (d $\sim$ 120 pc) only 2.8 pc apart, are studied and compared in order to explore the chemical structure of the earliest stages of star formation.
IRAS 16293-2422, located in L1689N, is a well-studied multiple system with the main components A and B separated by 800 AU.
Source A drives a complex outflow, has a large, close to edge-on, disk-like structure with a radius of 200 AU (Oya et al. 2016), and has a bolometric luminosity of 18 L$_{\odot}$ (Jacobsen et al. submitted).
Source B has a bolometric luminosity of 3L$_{\odot}$ (Jacobsen et al. submitted) and is positioned face-on, thus the presence of outflow and disk-like structure is difficult to determine.
VLA 1623-2417, located in L1688 ($\rho$ Oph A), is a triple protostellar system with a prominent outflow. The three components have similar inclinations, but are at different evolutionary stages. Source A, with a bolometric luminosity of 1 L$_{\odot}$, has a confirmed rotationally supported disk with a radius of 150--180 AU (Murillo et al. 2013). VLA1623-2417 has been shown to be largely line poor, in contrast to IRAS 16293-2422 which shows complex organic molecules.
This study does not focus on the compact ``hot core" region ($<$1"), but on the more extended cold and warm gas.

The results and analysis of Atacama Large Millimeter/submillimeter Array (ALMA) observations of \ce{DCO+}, \ce{N2D+}, \ce{c-C3H2} and \ce{C2H} toward both embedded protostellar systems are presented here.
\ce{DCO+} and \ce{N2D+} are good tracers of cold molecular gas where \ce{CO} is frozen out onto the dust grains, and are found in the envelope of young embedded protostars.
\ce{c-C3H2} and \ce{C2H} trace warmer gas due to being produced when material is UV-irradiated.
The two molecules are generally found in photon-dominated regions (PDRs) and outflow cavities in protostellar disk-like structures.

IRAS 16293-2422 was observed in the "Protostellar Interferometric Line Survey" (PILS) program (Project-ID: 2013.1.00278.S; PI: Jes K. J{\o}rgensen) in Cycle 2 with Band 7, using both the main 12-m array and the Atacama Compact Array (ACA). The combined observations provide a synthesized beam of 0.5". Detailed description of the observations and data reduction is given in J{\o}rgensen et al. (2016).

VLA 1623-2417 was observed in Cycle 0 and 2, centered on source A. Cycle 2 includes the ACA. Band 6 observations of \ce{DCO+} 3--2 provided a resolution of 0.85"$\times$0.65" in Cycle 0. The Band 6 observations of \ce{c-C3H2} have a resolution of 1.68"$\times$0.88". Band 7 observations of \ce{N2D+} have a spatial resolution of of 0.87"$\times$0.54".
Details on the data and reduction are given in Murillo et al. (2015; in prep.).
Additional single-dish Atacama Pathfinder EXperiment (APEX) observations of \ce{C2H} toward VLA 1623-2417 were carried out using single-pointing mode with a beam of 28.7" (Murillo et al. in prep.).

\section{Results and Analysis}
\ce{DCO+} 5--4 is detected in IRAS 16293-2422 around source A in a half-crescent shape, offset from the source position by 2.5" (Figure 1). 
No \ce{DCO+} emission is found around source B.
Toward VLA 1623-2417, \ce{DCO+} 3--2 is also found around source A, but not the other two components of the system, B and  W.
\ce{DCO+} 3--2 peaks about 2.5" away from source A to the south (Figure 1).
In order to understand the observed \ce{DCO+} emission, the simple chemical model from Murillo et al. (2015) is used, coupled with spherically symmetric density and temperature profiles for each source (IRAS 16293-2422: Crimier et al. 2010; VLA 1623-2417: J{\o}rgensen et al. 2002).
The model results show that the peak of the \ce{DCO+} emission is not where expected from the spherically symmetric profiles (IRAS 16293-2422: $\sim$10"; VLA 1623-2417: $\sim$5").
Instead, a drop in the temperature profile is needed to reproduce the observed \ce{DCO+} peak position. 
The model results indicate that the temperature at the \ce{DCO+} peak position is 18 K and 14 K for IRAS 16293-2422 and VLA 1623-2417, respectively.
This drop in temperature is most likely caused by the presence of the disk(-like) structure, which shadows the envelope material at its edge, generating an inward shift of cold molecules along the flattened structure but not elsewhere.

The position of the \ce{N2D+} 3--2 detected around IRAS 16293-2422 A (Jorgensen et al. 2011) is also well described by the temperature structure obtained from the \ce{DCO+} observations.
For VLA 1623-2417, \ce{N2D+} 3--2 is not detected with ALMA. 
Single-dish observations find \ce{N2D+} peaks to the north and away from VLA 1623-2417 by 60" (Punanova et al. 2016).
The lack of compact \ce{N2D+} emission in the inner envelope of VLA 1623-2417 can be explained as \ce{N2}, the precursor of \ce{N2D+}, being frozen onto the dust grains due to low gas and dust temperatures.
This is also in agreement with the temperature structure obtained from modeling the \ce{DCO+} observations.

\ce{c-C3H2} is found to trace the outflow cavity of VLA 1623-2417 A, and one side of the south outflow cavity of IRAS 16293-2422 A (Figure 1).
This is consistent with observations of other objects that find \ce{c-C3H2} in UV-irradiated regions.
No \ce{c-C3H2} is found in the disk(-like) structures of either source, in contrast to what is observed in L1527 (Sakai et al. 2014).
\ce{C2H} is detected on-source of VLA 1623-2417 A with single-dish observations, and is either located in the envelope or outflow cavity, indicating spatial correlation with the observed \ce{c-C3H2} emission around source A.
In contrast, toward IRAS 16293-2422 A, \ce{C2H} does not show any spatial correlation with \ce{c-C3H2} (Figure 1).
Instead, \ce{C2H} is located in a clumpy filament-like structure stretching from north to south, and apparently passing through IRAS 16293-2422 B.
This lack of correlation is surprising, since both molecules have been found to have a correlation from low- and high-mass protostars, as well as in diffuse clouds and PDRs.
A possible explanation for the lack of \ce{C2H} in the outflow cavity could be the reaction of \ce{C2H} with sulfur, nitrogen or carbon chains. 
However, molecules such as \ce{C2S} and \ce{C2N} are not found in the same region as \ce{c-C3H2}. 
Another explanation could be that \ce{C2H} is reacting with carbon chains, producing even larger carbon chains and molecular hydrogen, and thus cannot be detected due to the lack of a dipole moment.
The physical structure of IRAS 16293-2422 could also impact the spatial distribution of \ce{c-C3H2} and \ce{C2H}.

\section{Conclusions}
Using cold (\ce{DCO+}, \ce{N2D+}) and warm (\ce{c-C3H2}, \ce{C2H}) molecules, the chemical structure of two low-mass embedded protostars, IRAS 16293-2422 and VLA 1623-2417, is studied.
In IRAS 16293-2422 and VLA 1623-2417, the spatial distribution of \ce{DCO+}, \ce{N2D+} and \ce{c-C3H2} can be explained with the same physical structures: drop in temperature for the cold molecules, and irradiated outflow cavities for \ce{c-C3H2}.
\ce{C2H}, on the other hand, is spatially correlated in VLA 1623-2417 but not in IRAS 16293-2422.
While still a chemical puzzle, the anti-correlation of \ce{C2H} and \ce{c-C3H2} in IRAS 16293-2422 could be due to the complex structure of the system.
Two conclusions can be drawn from the comparison of these two systems.
\begin{enumerate}
	\item Temperature is an important controlling factor in the chemical structure of embedded protostars.
	\item The presence of disk(-like) structures produces a drastic impact on the physical and chemical structure of the protostar, since the temperature structure becomes asymmetric and cold close to the source.
\end{enumerate}
It is interesting to note that both systems studied here are multiple systems, and that the individual components have different chemical structures.  
This has also been observed in other embedded systems (e.g. NGC1333 SVS13: Chen et al. 2009; NGC1333 IRAS4A: Persson et al. 2012, L{\'o}pez-Sepulcre et al. 2017).
Thus, it is possible that there is no relation between multiplicity and chemical structure, at least in the embedded phase. 
In later stages, when the envelope has cleared, perhaps multiplicity plays a larger role in the chemical structure of stellar siblings.

\textit{Acknowledgments: The author thanks E. F. van Dishoeck, J. K. J{\o}rgensen, M. H. D. van der Wiel, C. Walsh, S. Bruderer,  D. Harsono,  M. N. Drozdovskaya, S.-P. Lai, C. M. Fuchs and H. Calcutt for useful discussions on the observaitons and modelling that made this work possible.}

\begin{figure}
	\centering
	\includegraphics[width=\textwidth]{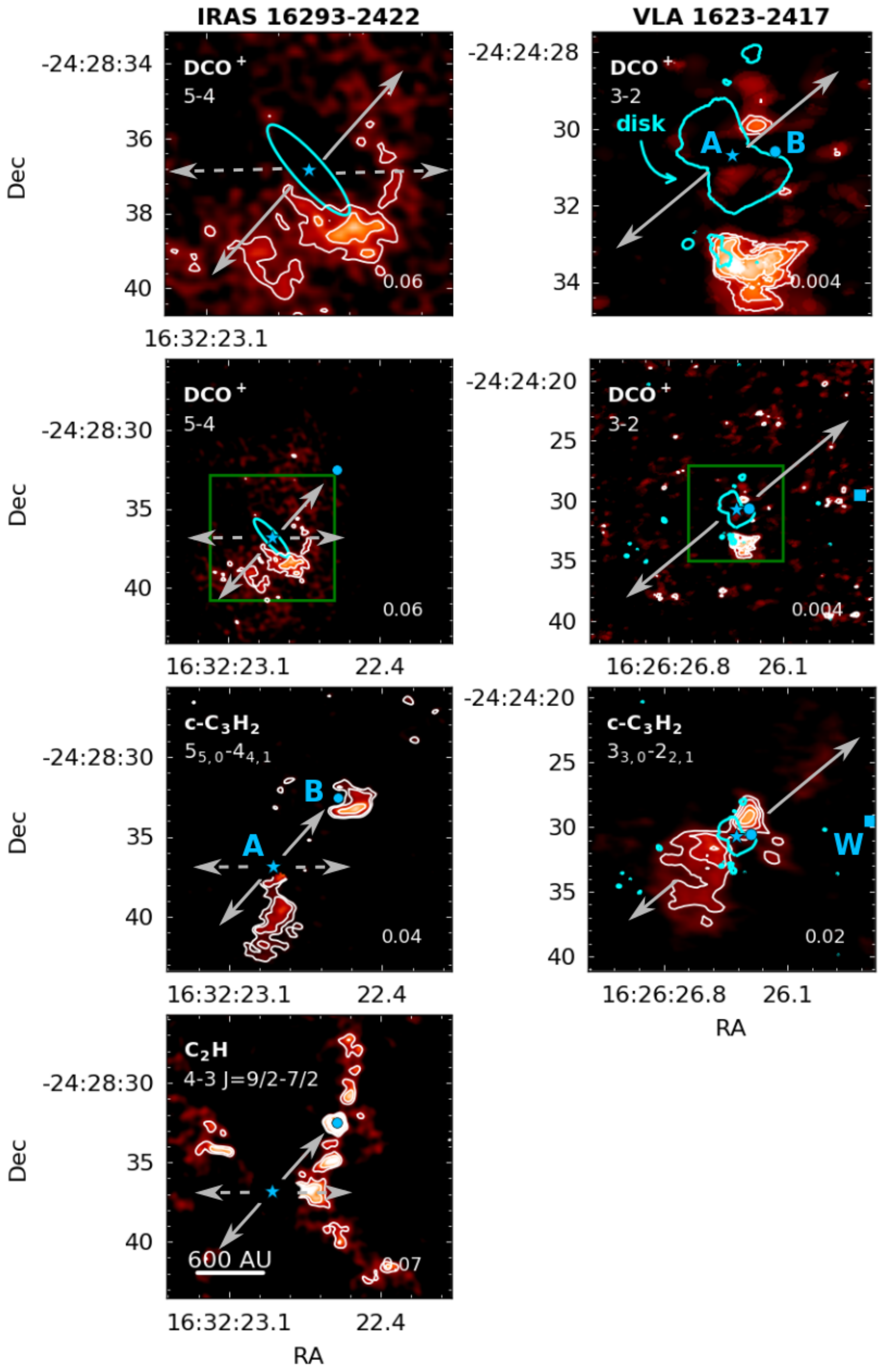}
	\caption{Intensity integrated maps of \ce{DCO+}, \ce{c-C3H2} and \ce{C2H} toward IRAS 16293-2422 (left column), and VLA 1623-2422 (right column). The green box shows the zoomed region shown in the top row. The gray arrows indicate the direction of the outflow(s). The cyan contour traces the extent of the observed disk(-like) structures in each system. Contours are in steps of 2, 3 and 5$\sigma$ for IRAS 16293-2422. For VLA 1623-2417, \ce{DCO+} is in steps of 4, 5 and 6$\sigma$, and \ce{c-C3H2} is in steps of 3, 4, 5 and 6$\sigma$, with $\sigma$ (Jy~beam$^{-1}$~km~s$^{-1}$) indicated in the lower right of each panel.}
\end{figure}


\begin{thebibliography}{}
\bibitem[Chen \etal\ (2009)]{chen2009}{Chen, X.; Launhardt, R. \& Henning, T, 2009, \textit{ApJ},a 691, 1729}

\bibitem[Crimier \etal\ (2010)]{crimier2010}{Crimier, N.; Ceccarelli, C.; Maret, S.; Bottinelli, S. \etal\, 2010, \textit{A\&A} 519, 65}

\bibitem[J{\o}rgensen \etal\ (2002)]{jorgensen2002}{J{\o}rgensen, J. K.; Sch{\"o}ier, F. L.; van Dishoeck, E. F., \textit{A\&A}, 389, 908}

\bibitem[J{\o}rgensen \etal\ (2011)]{jorgensen2011}{J{\o}rgensen, J. K.; Bourke, T. L.; Nguyen Luong, Q.; Takakuwa, S., \textit{A\&A}, 534, 100}	

\bibitem[J{\o}rgensen \etal\ (2016)]{jorgensen2016}{J{\o}rgensen, J. K.; van der Wiel, M. H. D.; Coutens, A.; Lykke, J. M. \etal\, 2016, \textit{A\&A}, 595, 117}

\bibitem[L{\'o}pez-Sepulcre \etal\ (2009)]{}{L{\'o}pez-Sepulcre, A.; Sakai, N.; Neri, R.; Imai, M. \etal\, 2017, \textit{arXiv:1707.03745}}

\bibitem[Murillo \etal\ (2013)]{murillo2013}{Murillo, N. M.; Lai, S.-P.; Bruderer, S.; Harsono, D. \& van Dishoeck, E. F., 2013, \textit{A\&A}, 560, 103}

\bibitem[Murillo \etal\ (2015)]{murillo2015}{Murillo, N. M., Bruderer, S., van Dishoeck, E. F., Walsh, C., \etal\, 2015, \textit{A\&A}, 579, 114}

\bibitem[Persson \etal\ (2012)]{persson2012}{Persson, M. V.; J{\o}rgensen, J. K. \& van Dishoeck, E. F., 2012, \textit{A\&A}, 541, 39}

\bibitem[Punanova \etal\ 2016]{punanova2016}{Punanova, A.; Caselli, P.; Pon, A.; Belloche, A. \& Andr{\'e}, Ph., 2016 \textit{A\&A}, 587, 118}

\bibitem[Oya \etal\ (2016)]{oya2016}{Oya, Y.; Sakai, N.; L{\'o}pez-Sepulcre, A.; Watanabe, Y.; Ceccarelli, C. \etal\, 2016, \textit{ApJ}, 824, 88}

\bibitem[Sakai \etal\ (2014)]{sakai2014}{Sakai, Nami; S., T.; Hirota, T.; Watanabe, Y.; Ceccarelli, C. \etal\, 2014, \textit{Nature}, 507, 78}


\end{thebibliography}
\end{document}